\documentclass[aps,prd,longbibliography ,twocolumn,nofootinbib,10pt]{revtex4-2}
\usepackage{graphicx}
\usepackage{amsmath}
\usepackage{amsfonts}
\usepackage{graphicx}
\usepackage{dsfont}
\usepackage[german, english]{babel}
\usepackage{amssymb}
\usepackage{ifthen}
\usepackage{ulem}
\usepackage{color}
\usepackage{float}
\usepackage{physics }
\usepackage{hyperref}
\usepackage{bbold}
\usepackage{lipsum}
\usepackage{nccmath}
\usepackage{xcolor}

\newcommand{\da}{^\dagger}

\graphicspath{{images/}}

\begin{document}

\author{Anton Talkachov}
\email{anttal@kth.se}
\author{Egor Babaev}
\affiliation{Department of Physics, KTH-Royal Institute of Technology, SE-10691, Stockholm, Sweden}

\title{Wave functions and edge states in rectangular honeycomb lattices revisited: nanoflakes, armchair and zigzag nanoribbons and nanotubes}

\begin{abstract}
Properties of bulk and boundaries of materials can, in general, be quite different, both for topological and non-topological reasons.
One of the simplest boundary problems to pose is the tight-binding problem of noninteracting electrons on a finite honeycomb lattice. Despite its simplicity, the problem is quite rich and directly related to the physics of graphene.
We revisit this long-studied problem and present an analytical derivation of the electron spectrum and wave functions for graphene rectangular derivatives.
We provide an exact analytical description of extended and localized states, the transition between them, and a special case of a localized state when the wave function is nonzero only at the edge sites.
The later state has zero energy, we discuss its existence in zigzag nanoribbons, zigzag nanotubes with number of sites along a zigzag edge divisible by 4, and rectangular graphene nanoflakes with an odd number of sites along both zigzag and armchair edges.
\end{abstract}

\maketitle

\section{Introduction}

The successive downscaling of graphene-based devices with atomic level precision \cite{cai2010atomically} shows the significant effect of edges on the electronic structure of graphene \cite{neto2009electronic}, which has experimental evidence \cite{kobayashi2005observation, kobayashi2006edge, tao2011spatially}.
 Insights into the electronic properties of graphene and its derivatives can be obtained from exact analytical approaches.
 There are two basic approaches for describing rectangular structures: based on the division of graphene into two sublattices \cite{wallace1947band, wakabayashi2012nanoscale, Wakabayashi_2010, wakabayashi1999electronic, moradinasab2012analytical, Saroka2017Optics, yorikawa2021edge} (equivalently, choosing two atom unit cell) or choosing a unit cell consisting of four atoms \cite{malysheva2008spectrum, Malysheva2017Analytic, Onipko2018Revisit, onipko2008spectrum, ruseckas2011spectrum}.
The first one is usually applied when describing infinite systems and nanoribbons, the second one is used mostly for finite systems.
There are also two basic edge shapes for nanoribbons: armchair and zigzag.
It has been shown that zigzag nanoribbons possess localized edge states with energies close to the Fermi level \cite{nakada1996edge, fujita1996peculiar, wakabayashi1999electronic, Wakabayashi_2010, wakabayashi2012nanoscale, Saroka2017Optics}.
The edge states have been predicted to be important in transport \cite{luck2014unusual}, electromagnetic \cite{brey2007elementary}, and optical properties \cite{Saroka2017Optics, hsu2007selection}.
In contrast, edge states are absent for armchair nanoribbons \cite{wakabayashi2012nanoscale, Wakabayashi_2010, neto2009electronic}.
The same results have been shown from the topological point of view \cite{ryu2002topological}.
Besides graphene, the problem is also relevant for artificial honeycomb structure and ultracold atoms on honeycomb optical lattices.

We revisit the basic problem of $\pi$-electrons in rectangular graphene geometries (armchair and zigzag nanoribbons, nanotubes, finite samples). 
We establish the transition points between extended and localized states.
It is often assumed that these points can be treated with wave functions for extended waves.
However, we show that the direct approach leads to a trivial (zero) wave function everywhere and we need to take the limit to these points from left and right.
We discuss that at these transitions, wave functions have a linear dependence on the site index.
We also describe the two entirely localized zero energy states with nonzero wave functions only at the edge sites for zigzag nanoribbons, zigzag nanotubes, and one geometry of rectangular graphene nanoflake, that to the best of our knowledge were not reported before.
We discuss wave functions in a rectangular graphene system including expressions for localized states.

This paper is organized as follows. In Sec. \ref{sec:Infinite structure}, we recap the eigenvalue problem for the infinite honeycomb lattice, where wave functions are running waves in both $x$ and $y$ directions.
In Sec. \ref{sec:Systems finite in one direction}, we switch to problems for finite system size in one direction: armchair and zigzag nanoribbons and nanotubes. Here we show that edge states are realizable only for nanoribbons and nanotubes with zigzag open edges.
Importantly, there is an entirely localized edge state with zero energy, where wave functions are nonzero only at the edge sites.
In Sec. \ref{sec:Finite sample}, we find wave functions for extended and localized states in a finite sample with rectangular geometry.
We show analytically that there is only one possible geometry which allows for an entirely localized edge state.

\section{The infinite system} \label{sec:Infinite structure}

\begin{figure}
    \centering
    \includegraphics[width=0.9\linewidth]{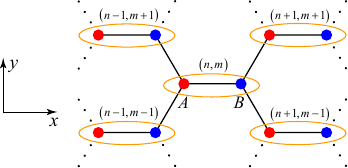}
    \caption{The honeycomb lattice in real space, where the red (blue) circles mean an $A$ ($B$)-sublattice site.}
    \label{fig: honeycomb infinite structure}
\end{figure}

Let us first look at the infinite honeycomb structure made from identical atoms.
We divide these atoms into two groups $(A, B)$ which form triangular lattices (Fig. \ref{fig: honeycomb infinite structure}).
The effective Hamiltonian for the system reads
\begin{equation}
\label{eq:effective_Hamiltonian}
    H_\text{eff} = -t\sum_{\langle \textbf{i},\textbf{j} \rangle} \left( a_{\textbf{i}}\da b_{\textbf{j}} + b_{\textbf{j}}\da a_{\textbf{i}} \right).
\end{equation}
Here, $a_{\textbf{i}}\da (a_{\textbf{i}})$ is the creation (annihilation) operator for an electron on site $A$ in the cell whose position is described by the vector $\textbf{i}=(n, m)$, where $n$ ($m$) specifies horizontal (vertical) position with indexation illustrated in Fig. \ref{fig: honeycomb infinite structure}.
The same applies to operators $b_{\textbf{i}}\da$ and $b_{\textbf{i}}$ which correspond to sites $B$.
This Hamiltonian describes kinetic energy (hopping between nearest-neighbor sites $\langle \textbf{i},\textbf{j} \rangle$ without spin flip), parameterized by the hopping integral $t$ ($t>0$ and assumed to be constant in space).
A general state can be written as
\begin{equation}
    \lvert \Psi \rangle = \sum_{\textbf{i}} \left( \psi_{A,\textbf{i}} a_{\textbf{i}}\da + \psi_{B,\textbf{i}} b_{\textbf{i}}\da \right) \lvert 0 \rangle,
\end{equation}
where $\psi_{A,\textbf{i}}$ ($\psi_{B,\textbf{i}}$) is the real space wave function describing an electron on the $A$ ($B$) sublattice, $\lvert 0 \rangle$ denotes the vacuum state with no particle present.
We will assume the plane wave form of the wave functions
\begin{equation}
\label{eq:wave function definition}
    \mqty(\psi_{A,\textbf{i}} \\ \psi_{B,\textbf{i}}) = e^{i (k_x n + k_y m)} \mqty(f_A(\textbf{k}) \\f_B(\textbf{k})),
\end{equation}
we come to the Schr$\ddot{\text{o}}$dinger equation
\begin{equation}
\label{eq:Schrodinger}
    H_\text{eff} \lvert \Psi \rangle = E \lvert \Psi \rangle
\end{equation}
which takes the form
\begin{equation}
\begin{gathered}
\begin{pmatrix}
    0 && -t(1 + 2 e^{-i k_x} \cos{k_y}) \\
    -t(1 + 2 e^{i k_x} \cos{k_y}) && 0            
\end{pmatrix}
    \mqty(f_A(\textbf{k}) \\f_B(\textbf{k})) \\
    = E \mqty(f_A(\textbf{k}) \\f_B(\textbf{k})).   
\end{gathered}
\end{equation}  
Eigenvalues of the matrix in this equation give possible energies:
\begin{equation}
\label{eq:energies}
\begin{gathered}
    E_s = s \cdot t \cdot \epsilon (k_x, k_y),
    \\
    \epsilon (k_x, k_y) = \sqrt{3 + 4 \cos{k_x} \cos{k_y} + 2 \cos{2 k_y}},
\end{gathered}
\end{equation}
where we have introduced the parameter $s = \pm 1$ to distinguish between valence ($-1$) and conduction ($+1$) bands. The corresponding eigenvectors $(f_A(\textbf{k}),f_B(\textbf{k}))^T$ allow us to calculate normalized wave functions (\ref{eq:wave function definition}) for the infinite honeycomb lattice:
\begin{equation}
\label{eq:wave functions infinite system}
    \mqty(\psi_{A,\textbf{i}} \\ \psi_{B,\textbf{i}}) = \frac{1}{\sqrt{2}} e^{i (k_x n + k_y m)} \mqty(- \frac{s(1 + 2 e^{-i k_x} \cos{k_y})}{\epsilon (k_x, k_y)}
    \\
          1).
\end{equation}
Note that $\psi_{A,\textbf{i}}$ and $\psi_{B,\textbf{i}}$ is only a basis. General wave functions that describe the system are superpositions of all these possible states.

\section{Systems finite in one direction} \label{sec:Systems finite in one direction}

\subsection{Wave functions for armchair nanoribbons and nanotubes} \label{sec:wave functions for armchair nanoribbons}

An armchair nanoribbon is a sample of the honeycomb lattice arranged like in Fig. \ref{fig: armchair nanoribbon}, which is finite in $y$. Wave functions, in this case, can be obtained as a superposition of travelling waves in $+y$ and $-y$ directions:

\begin{figure}
    \centering
    \includegraphics[width=0.99\linewidth]{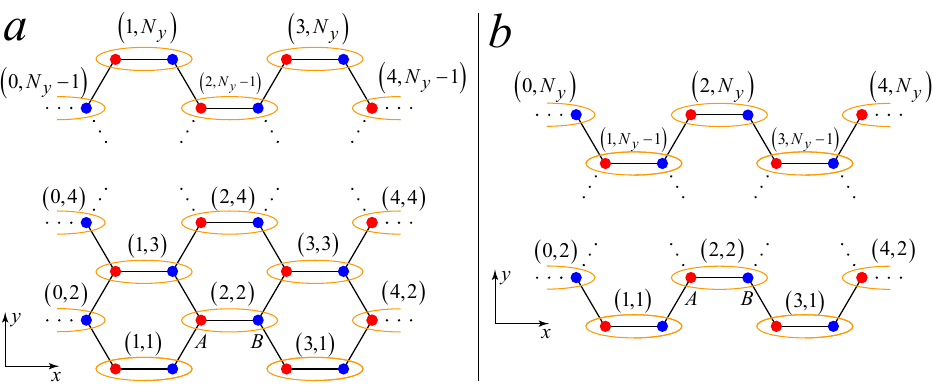}
    \caption{Honeycomb lattice with armchair edges (armchair nanoribbon) in two different cases: (a) $N_y$ is odd, (b) $N_y$ is even.}
    \label{fig: armchair nanoribbon}
\end{figure}

\begin{equation}
\begin{gathered}
    \mqty(\psi_{A,\textbf{i}} \\ \psi_{B,\textbf{i}} ) = \Bigl[ c_1 \frac{1}{\sqrt{2}}  \mqty(- \frac{s(1 + 2 e^{-i k_x} \cos{k_y})}{\epsilon (k_x, k_y)} 
    \\
          1) e^{i k_y m}\\
    + c_2 \frac{1}{\sqrt{2}}  \mqty(- \frac{s(1 + 2 e^{-i k_x} \cos{(-k_y)})}{\epsilon (k_x, -k_y)}
    \\
          1) e^{i (-k_y) m} \Bigr]  e^{i k_x n}
    \\
    = \frac{1}{\sqrt{2}}  \mqty(-\frac{s(1 + 2 e^{-i k_x} \cos{k_y})}{\epsilon (k_x, k_y)}
    \\
          1)
    \left[ c_1 e^{i k_y m} + c_2e^{- i k_y m}
    \right] e^{i k_x n},
\end{gathered}
\end{equation}
where $c_1$ and $c_2$ are normalization constants. These wave functions should satisfy open boundary conditions (some authors \cite{fujita1996peculiar} say that the dangling bonds at the edge are terminated by hydrogen atoms so they do not contribute to the electronic states):
\begin{equation}
\label{eq: BC for armchair}
    \mqty(\psi_{A,\textbf{i} = (n, 0)} \\ \psi_{B,\textbf{i} = (n, 0)}) = \mqty(0 \\ 0),
    \quad
    \mqty(\psi_{A,\textbf{i} = (n, N_y +1)} \\ \psi_{B,\textbf{i} = (n, N_y +1)}) = \mqty(0 \\ 0).
\end{equation}
Physically, this means that electrons are absent outside the considered system (their wave function equals zero).
This leads to the following relation between $c_1$ and $c_2$, and to the possible values of wavenumber $k_y$:
\begin{equation}
\label{eq:ky}
    c_2 = - c_1, \qquad k_y = \frac{\pi j_y}{N_y + 1}, \quad j_y = 1, 2, 3, \ldots N_y.
\end{equation}

Wavenumber $k_x$ should be treated as a continuous variable in the case of an infinite honeycomb sheet in the $x$ direction.
We consider a case of a honeycomb nanotube with open armchair boundaries ($N_x$ has to be even).
For this case we use periodic boundary conditions along the $x$ axis: $\psi_{A,\textbf{i} = (n, m)} = \psi_{A,\textbf{i}' = (n + N_x, m)}$, $\psi_{B,\textbf{i} = (n, m)} = \psi_{B,\textbf{i}' = (n + N_x, m)}$.
This leads to a discrete set of $k_x$ values:
\begin{equation}
\label{eq:kx}
    k_x = \frac{2 \pi}{N_x} \nu_x, \quad \nu_x = 0,1,2, \ldots \frac{N_x}{2} - 1.
\end{equation}

A system consists of $N_x N_y$ sites, so the eigenvalue problem (\ref{eq:Schrodinger}) should have $N_x N_y$ different eigenvalues and eigenvectors. We have $N_y$ values of $k_y$ (\ref{eq:ky}), $N_x / 2$ values of $k_x$ (\ref{eq:kx}) and two different values of the parameter $s = \pm 1$.
The coefficient $c_1$ can be found from the following normalization conditions:
\begin{equation}
\label{eq:normalization}
\begin{gathered}
    \sum_{\textbf{i}} (\psi_{A,\textbf{i}}, \psi_{B,\textbf{i}}) (\psi_{A,\textbf{i}}, \psi_{B,\textbf{i}})\da = 1, \\
    \sum_{s, k_x, k_y} (\psi_{A,\textbf{i}}, \psi_{B,\textbf{i}}) (\psi_{A,\textbf{i}}, \psi_{B,\textbf{i}})\da = 1.    
\end{gathered}
\end{equation}
In the article, we always use the first of the rules to find the normalization coefficients, but the second one is also satisfied in all considered cases.
The constant $c_1$ has the form
\begin{equation}
    c_1 = - i \left( 2 N_x \cdot S(k_y, N_y) \right)^{-1/2},
\end{equation}
\begin{equation}
\label{eq:sum sin^2}
    S(k, N) =  \sum_{m=1}^{N} \sin^2 k m = \frac{1}{2} \left( N - \frac{\sin{k N}}{\sin{k}} \cos{k (N + 1)} \right),
\end{equation}
where we have introduced auxiliary function $S(k, N)$.
Note that this normalization coefficient is defined up to an arbitrary factor $e^{i \phi}$,
which is why wave functions from articles \cite{Wakabayashi_2010, wakabayashi2012nanoscale, Onipko2018Revisit, zheng2007analytical}
may look different from our final result:
\begin{equation} \label{eq: wave function armchair}
    \mqty(\psi_{A,\textbf{i}} \\ \psi_{B,\textbf{i}} ) = \frac{e^{i k_x n} \sin{k_y m}}{\sqrt{N_x \cdot S(k_y, N_y)}}  \mqty(-\frac{s(1 + 2 e^{-i k_x} \cos{k_y})}{\epsilon (k_x, k_y)}
    \\
          1).
\end{equation}
After renormalization, the results become identical.
These states are called extended because they describe waves which extend over the whole ribbon width.
Our way of presenting result is better for computer-based calculations, in the comparison with using square roots from complex numbers \cite{Wakabayashi_2010,wakabayashi2012nanoscale} which are multivalued functions.

There is a zero energy state for armchair nanoribbons of width $N_y = 3r - 1$ (where $r = 1, 2, 3, \ldots$) at $k_x = 0$, $k_y = 2 \pi / 3$.
The system with such a width in the $y$ direction is called metallic \cite{fujita1996peculiar, nakada1996edge}.
The wave function for $A$ sites is ill-defined at the point, because there exists the following $\frac{0}{0}$ uncertainty:
\begin{equation}
\begin{gathered}
    \lim_{k_x \rightarrow 0, k_y \rightarrow 2 \pi / 3} \frac{1 + 2 e^{-i k_x} \cos{k_y}}{\epsilon (k_x, k_y)} \\
    = \lim_{k_x \rightarrow 0, k_y \rightarrow 2 \pi / 3} \frac{i k_x - \sqrt{3} (k_y - 2 \pi /3)}{\sqrt{k_x^2 + 3 (k_y - 2 \pi /3)^2}} = e^{i \varphi},  
\end{gathered}
\end{equation}
where we have introduced the parameter $\varphi \in [0; 2 \pi)$ which depends on the ratio $k_x / (k_y - 2 \pi / 3)$.
The possible values of wavenumber $k_y$ (\ref{eq:ky}) are obtained from the boundary conditions.
The wavenumber $k_x$ is either a continuous variable for an infinite nanoribbon or is quantized (\ref{eq:kx}) due to periodic boundary conditions for armchair nanotubes.
One can see that $k_x$ and $k_y$ are independent consequently, $\varphi$ is arbitrary and can be chosen artificially.
It is convenient to choose $\varphi = \pi$, then wave functions at the point $k_x = 0$, $k_y = 2 \pi / 3$ have the form
\begin{equation} \label{eq: wave function armchair}
    \mqty(\psi_{A,\textbf{i}} \\ \psi_{B,\textbf{i}} ) = \frac{\sin{(2 \pi m / 3)}}{\sqrt{N_x \cdot S(2 \pi / 3, N_y)}}  \mqty(s  \\    1).
\end{equation}

\subsection{Wave functions for zigzag nanoribbons and nanotubes} \label{sec:wave functions for zigzag nanoribbons}
\begin{figure}
    \centering
    \includegraphics[width=0.99\linewidth]{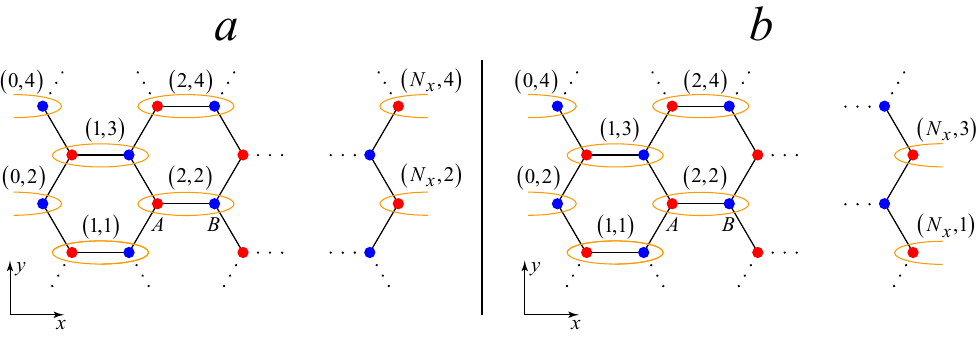}
    \caption{Honeycomb lattice with zigzag edges (zigzag nanoribbon) in two different cases: (a) $N_x$ is even, (b) $N_x$ is odd.}
    \label{fig: zigzag nanoribbon}
\end{figure}

Zigzag nanoribbons are systems that are finite in the $x$ direction but can be infinite in the $y$ direction (Fig. \ref{fig: zigzag nanoribbon}).

Let us apply an approach similar to the previous subsection:
obtain wave functions as a superposition of waves travelling in $\pm x$ directions:
\begin{equation}
\label{eq:zigzag_wv_initial}
\begin{gathered}
    \mqty(\psi_{A,\textbf{i}} \\ \psi_{B,\textbf{i}} ) = \Bigl[ c_3 \frac{1}{\sqrt{2}}  \mqty(- \frac{s(1 + 2 e^{-i k_x} \cos{k_y})}{\epsilon (k_x, k_y)} 
    \\
          1) e^{i k_x n}\\
    + c_4 \frac{1}{\sqrt{2}}  \mqty(- \frac{s(1 + 2 e^{-i (-k_x)} \cos{k_y})}{\epsilon (-k_x, k_y)}
    \\
          1) e^{i (-k_x) n} \Bigr]  e^{i k_y m}
    \\
    = \frac{1}{\sqrt{2}} \Bigl[ c_3 \mqty(- \frac{s(1 + 2 e^{-i k_x} \cos{k_y})}{\epsilon (k_x, k_y)} 
    \\
          1) e^{i k_x n}\\
    + c_4 \mqty(- \frac{s(1 + 2 e^{i k_x} \cos{k_y})}{\epsilon (k_x, k_y)}
    \\
          1) e^{-i k_x n} \Bigr]  e^{i k_y m},
\end{gathered} 
\end{equation}
where $c_3$ and $c_4$ are coefficients that can be found from normalization conditions (\ref{eq:normalization}). Wave functions (\ref{eq:zigzag_wv_initial}) satisfy the following boundary conditions:
\begin{equation}
\label{eq: BC for zigzag}
    \psi_{A,\textbf{i} = (0, m)} = 0,   \qquad   \psi_{B,\textbf{i} = (N_x, m)} = 0.
\end{equation}
These constraints lead to
\begin{equation}
\label{eq:numerical equation}
    \sin{k_x N_x} + 2 \cos{k_y} \sin{\left(k_x (N_x + 1) \right)} = 0,
\end{equation}
\begin{equation}
    c_2 = - c_1 e^{2 i k_x N_x}.
\end{equation}

Possible values for wavenumber $k_y$ are obtained using the assumption of periodic boundary conditions in $y$ direction (honeycomb nanotube with open zigzag edges): $\psi_{A,\textbf{i} = (n, m)} = \psi_{A,\textbf{i}' = (n, m + N_y)}$, $\psi_{B,\textbf{i} = (n, m)} = \psi_{B,\textbf{i}' = (n, m + N_y)}$.
In this case, the number of sites in the $y$ direction ($N_y$) has to be even.
The wavenumber satisfies
\begin{equation}
\label{eq:zGNR ky}
    k_y = \frac{2 \pi}{N_y} \nu_y, \quad \nu_y = 0, 1, 2, \ldots \frac{N_y}{2} - 1.
\end{equation}

The transcendental Eq. (\ref{eq:numerical equation}) contains $N_x$ different roots for the wavenumber $k_x$.
This equation has $N_x$ real nontrivial roots in the region $k_x \in \left( 0; \pi \right)$ for $k_y \in \left[ 0 ; k_y^c \right) \cup \left( \pi - k_y^c ; \pi \right]$ and $N_x - 1$ real roots otherwise, where
\begin{equation}
    k_y^c = \arccos\frac{N_x}{2 (N_x + 1)}.
\end{equation}
The other one in the latter case is a complex solution.
They can be obtained numerically, analytically using approximate formulas \cite{Malysheva2017Analytic} or using fitting formulas for roots \cite{moradinasab2012analytical} that are obtained from numerical results.
Values of the wavenumber $k_x = 0$ and $\pm \pi$ are called unphysical \cite{Wakabayashi_2010}, because they lead to a trivial wave function which equals zero at all sites.
The equation is similar to one analysed by Klein applied to graphene nanoribbons with an additional methylene group at every edge site almost 30 years ago \cite{klein1994graphitic}, so this problem has been known for a long time.

\subsubsection{Case $k_y \in \left[ 0 ; k_y^c \right) \cup \left( \pi - k_y^c ; \pi \right]$}

In this case, all roots are real and can be found numerically from Eq. (\ref{eq:numerical equation}). Corresponding wave functions describe extended states:
\begin{equation}
\label{eq:zGNR extended wave function}
    \mqty(\psi_{A,\textbf{i}} \\ \psi_{B,\textbf{i}} ) = - \sqrt{2} i c_3 e^{i (k_x N_x + k_y m)} \mqty(s_1 s \sin (k_x n)  \\
    \sin (k_x (N_x-n))),
\end{equation}
\begin{equation}
\label{eq:s1}
    s_1 = \medmath{\frac{ \sqrt{\sin^2 k_x} }{\sin k_x} \cdot \frac{ \sin (k_x (N_x+1))}{ \sqrt{\sin^2 (k_x (N_x+1)) }}} = \text{sign} \left(\sin (k_x (N_x+1)) \right).
\end{equation}
The sign function $s_1$ can be simplified if one introduces special indexation for $k_x$ roots to distinguish different subbands \cite{wakabayashi2012nanoscale, Saroka2017Optics}.
The normalization constant $c_3$ can be written in the following way
\begin{equation}
    c_3 = i e^{-i k_x N_x} \left( 2 N_y \cdot S(k_x, N_x) \right)^{-1/2},
\end{equation}
where the complex factor is chosen in this form to have a compact presentation for the wave functions:
\begin{equation}
\label{eq:zGNR extended wave function final}
    \mqty(\psi_{A,\textbf{i}} \\ \psi_{B,\textbf{i}} ) = \frac{e^{i k_y m}}{\sqrt{N_y \cdot S(k_x, N_x)}} \mqty(s_1 s \sin (k_x n)  \\
    \sin (k_x (N_x-n))).
\end{equation}
Similar results were obtained in papers \cite{Wakabayashi_2010, wakabayashi2012nanoscale, Saroka2017Optics, moradinasab2012analytical}.
However, there is no sign function depending on $k_x$ in articles \cite{Wakabayashi_2010, moradinasab2012analytical}.
The authors of paper \cite{Wakabayashi_2010} corrected the mistake in a later article \cite{wakabayashi2012nanoscale} where, due to special numeration of roots of Eq. (\ref{eq:numerical equation}) $k_x^r$, a factor of $(-1)^r$ appears in wave functions which play the same role as the function $s_1$ (\ref{eq:s1}) in our representation.
Their index $r$ is related to the band number in \cite{wakabayashi2012nanoscale}.
In the paper \cite{Saroka2017Optics} wave functions were derived in the limit $k_y \rightarrow 0$. They coincide with our wave functions (\ref{eq:zGNR extended wave function final}) for $k_y = 0$.

Existence of the sign function $s_1$ (or similarly factor $(-1)^r$ to divide subbands) reflects inversion symmetry of the electron wave function and plays an important role for the optical selection rules \cite{Saroka2017Optics, chung2011exploration}.

\subsubsection{Case $k_y \in \left(k_y^c; \frac{\pi}{2} \right) \cup \left( \frac{\pi}{2} ; \pi - k_y^c \right)$}
In this case $N_x - 1$ real roots for $k_x$ can be found from Eq. (\ref{eq:numerical equation}) and the corresponding wave functions have the form (\ref{eq:zGNR extended wave function final}).
One more root can be obtained by analytical continuation to the complex plane \cite{Wakabayashi_2010}:
\begin{equation} \label{eq:complex plane continuation}
    k_x \rightarrow \begin{cases}
                        \pi \pm i k_x^ {'}, & k_y \in \left( k_y^c ; \frac{\pi}{2} \right), \\
                        0 \pm i k_x^ {'}, & k_y \in \left( \frac{\pi}{2} ; \pi - k_y^c \right),
                    \end{cases}
\end{equation}
and describes the edge state which corresponds to wave functions localized in space.
For $k_y = \frac{\pi}{2}$ we need to obtain solutions separately and we will do this later.
With the ansatz (\ref{eq:complex plane continuation}), the function $\epsilon (k_x, k_y)$ inside eigenenergies $E_s$ (\ref{eq:energies}) and transcendental Eq. (\ref{eq:numerical equation}) transform to
\begin{equation}
\label{eq:special case energies}
\medmath{ \epsilon (k_x^ {'}, k_y) =
\begin{cases}
    \sqrt{3 - 4 \cosh{k_x^ {'}} \cos{k_y} + 2 \cos{2 k_y}}, & k_y \in \left( k_y^c ; \frac{\pi}{2} \right), \\
    \sqrt{3 + 4 \cosh{k_x^ {'}} \cos{k_y} + 2 \cos{2 k_y}}, & k_y \in \left( \frac{\pi}{2} ; \pi - k_y^c \right),
\end{cases}
}
\end{equation}
\begin{equation}
\label{eq:hyperbolic equations}
\medmath{
\begin{cases}
    \sinh{k_x^ {'} N_x} - 2 \cos{k_y} \sinh{\left(k_x^ {'} (N_x + 1) \right)} = 0, & k_y \in \left( k_y^c ; \frac{\pi}{2} \right), \\
    \sinh{k_x^ {'} N_x} + 2 \cos{k_y} \sinh{\left(k_x^ {'} (N_x + 1) \right)} = 0, & k_y \in \left( \frac{\pi}{2} ; \pi - k_y^c \right).
\end{cases}
}
\end{equation}
Each of Eq. (\ref{eq:hyperbolic equations}) contains two opposite roots.
But after substituting them to find wave functions and normalizing, one can see that these roots describe identical wave functions with coinciding energies.
Therefore we are looking for only one (positive) root which is to be found numerically.
Both of Eq. (\ref{eq:hyperbolic equations}) can be joined to one \cite{Wakabayashi_2010} which has form
\begin{equation}
    \sinh{k_x^ {'} N_x} - 2 \lvert \cos{k_y} \rvert \sinh{\left(k_x^ {'} (N_x + 1) \right)} = 0.
\end{equation}

We need to obtain wave functions after finding all values of $k_y$ from (\ref{eq:zGNR ky}), the corresponding $k_x$ and energies.
Related wave functions that describe complex solutions for $k_y \in \left( \frac{\pi}{2} ; \pi - k_y^c \right)$ do not contain a sign functions, but there is a sign function $(-1)^n$ depending on horizontal position for $k_y \in \left( k_y^c ; \frac{\pi}{2} \right)$:
\begin{equation}
\label{eq:zGNR localized wave function}
\begin{gathered}
    \mqty(\psi_{A,\textbf{i}} \\ \psi_{B,\textbf{i}} ) = \sqrt{2} c_3^ {'} e^{-k_x^ {'}  N_x+i k_y m} \mqty(s \sinh (k_x^ {'}  n)     \\      \sinh (k_x^ {'}  (N_x-n)) ) \\
    \cdot 
    \begin{cases}
        (-1)^n, & k_y \in \left( k_y^c ; \frac{\pi}{2} \right), \\
        1, & k_y \in \left( \frac{\pi}{2} ; \pi - k_y^c \right).
    \end{cases}
\end{gathered}
\end{equation}
Their derivation is located in Appendix \ref{app:Derivation of wave functions for localized case}.
It is convenient to present the normalization constant $c_3^{'}$ in the following way:
\begin{equation}
    c_3^{'} = e^{k_x^ {'}  N_x} \left( 2 N_y \cdot S_{\text{hyp}}(k_x^{'}, N_x) \right)^{-1/2},
\end{equation}
where
\begin{equation}
\label{eq:hyperbolic sum}
\begin{gathered}
    S_{\text{hyp}}(k, N) = \sum_{m=1}^{N} \sinh^2 k m \\
    = \frac{1}{4} \left(\frac{\sinh{k (2 N + 1)}}{\sinh{k}} - (2 N + 1) \right).    
\end{gathered}
\end{equation}
wave functions (\ref{eq:zGNR localized wave function}) can be finally written as
\begin{equation}
\label{eq:zGNR localized wave function final}
\begin{gathered}
    \mqty(\psi_{A,\textbf{i}} \\ \psi_{B,\textbf{i}} ) = \frac{e^{i k_y m}}{\sqrt{N_y \cdot S_{\text{hyp}}(k_x^{'}, N_x)}} \mqty(s \sinh (k_x^ {'}  n)     \\      \sinh (k_x^ {'}  (N_x-n)) ) \\
    \cdot 
    \begin{cases}
        (-1)^n, & k_y \in \left( k_y^c ; \frac{\pi}{2} \right), \\
        1, & k_y \in \left( \frac{\pi}{2} ; \pi - k_y^c \right).
    \end{cases}
\end{gathered}
\end{equation}
Comparing the result to others one can see that in paper \cite{Wakabayashi_2010} for the case $k_y \in \left( k_y^c ; \frac{\pi}{2} \right)$ in some formulas (Eq.(B.40), (23) in referring article) the sign function exist in the form $e^{i \pi n}$ (converting to our type of numeration), but in some formulas (B.44) it is absent.
Factor $e^{i \pi n}$ in \cite{Wakabayashi_2010} plays the same role as factor $(-1)^n$ in our Eq. (\ref{eq:zGNR localized wave function final}).
For the case $k_y \in \left( \frac{\pi}{2} ; \pi - k_y^c \right)$ they also obtained result without sign function.

In the Ref. \cite{wakabayashi2012nanoscale}, authors work with $k_y$ in the range $(-\pi/2 ; \pi/2]$.
They have continuation to the complex plane of one type ($k_x = \pi \pm i k_x^ {'}$) and it is right for their choice of $k_y$.
However, they obtained localized wave functions with extra sign function which depend on indexation of $k_x$ roots (as was explained in the previous subsection), which is incorrect (the vanishing procedure is described in Appendix \ref{app:Derivation of wave functions for localized case}).

In the article \cite{Saroka2017Optics} the range  and possible values of $k_y$ are not specified, but from figures one can see that they used $k_y \in \left( -\pi/2; \pi/2 \right]$ (in our notation).
They used the transfer matrix method and in the step when introducing new variable at Eq. (11), they chose a sign that lead to a changing sign of one term in the energy (14) and transcendental Eq. (19).
Further, they chose the continuation of $k_x = 0 \pm i k_x^ {'}$ which is correct for their equation type.
They obtained a result identical to our Eq. (\ref{eq:zGNR localized wave function final}) for $k_y \in \left( \frac{\pi}{2} ; \pi - k_y^c \right)$ with the same type of continuation.

\subsubsection{Case $k_y = \pi / 2$} \label{sec:Case k_y = pi/2}
To the best of our knowledge previous papers dedicated to analytical calculations of localized wave functions in the honeycomb lattice, e.g. \cite{Saroka2017Optics, Wakabayashi_2010} do not discuss this special value of $k_y$ (or write that this value is included in the range of the previous subsection and wave functions should be treated with formulas (\ref{eq:zGNR localized wave function final}) \cite{wakabayashi2012nanoscale}).
Below we will show that at this point wave functions should be derived more carefully.

For this value of wavenumber $k_y$, Eq. (\ref{eq:numerical equation}) simplifies to
\begin{equation} \label{eq:sin kxNx=0}
    \sin{k_x N_x} = 0,
\end{equation}
leading to $N_x - 1$ real roots which can be found analytically:
\begin{equation}
    k_x = \frac{\pi}{N_x} j_x, \quad j_x = 1, 2, 3, \ldots N_x - 1.
\end{equation}

Like in previous cases we expect to have one eigenvalue that describes a localized state because on the $k_y$ axis this case is bounded from left and right by localized type of solutions for the eigenvalue problem.
Straightforward method for finding solution $k_x$ in complex plane leads to trivial nonphysical results $k_x = 0, \pi$.
We find solution for this case as a limit for the previous subsection's results.

First, we consider the left-hand limit: $k_y = \frac{\pi}{2} - \eta$, where $\eta \rightarrow 0^+$.
In this case the complex plane continuation has the form $k_x = \pi \pm i k_x^ {'}$. There is a numerically obtained tendency: decreasing $\eta$ leads to increasing the imaginary part of $k_x$.
The eigenvalue equation (first of Eq.(\ref{eq:hyperbolic equations})) transforms to
\begin{equation}
    \frac{1}{2} e^{k_x^ {'} N_x} - 2 \cos{\left( \frac{\pi}{2} - \eta \right)} \cdot \frac{1}{2} e^{k_x^ {'} (N_x + 1)} = 0,
\end{equation}
which is obtained in the limit $\eta \rightarrow 0$ ($k_x^ {'} \rightarrow \infty$). This has one solution
\begin{equation}
    k_x^ {'} = - \ln 2 \eta,
\end{equation}
which tends to infinity when $\eta \rightarrow 0$ what confirms our numerical results.

The function $\epsilon (k_x^ {'}, \frac{\pi}{2})$ (\ref{eq:special case energies}) inside eigenenergies $E_s$ now can be written as
\begin{equation}
\label{eq:zero energy}
\begin{gathered}
   \epsilon (k_x^ {'}, \frac{\pi}{2}) = \sqrt{3 - 4 \cdot \frac{1}{2} e^{k_x^ {'}} \cos{k_y} + 2 \cos{2 k_y}} \\
    = \lim_{\eta \rightarrow 0} \sqrt{3 - 2 e^{-\ln{2 \eta}} \cos{(\frac{\pi}{2} - \eta)} + 2 \cos{2 (\frac{\pi}{2} - \eta)}} \\
    = \lim_{\eta \rightarrow 0} \sqrt{3 - 2 \cdot (2 \eta)^{-1} \eta + 2 \cdot (-1)} = 0.
\end{gathered}
\end{equation}
Therefore, one of the wavenumbers $k_x$ at this special point $k_y = \pi / 2$ describes a zero energy state.
This state has double degeneracy, because there are two different wave functions previously related to valence and conduction bands, describing the one energy state.
Before finding the explicit form of the wave functions in this case, let us first simplify the summation function (\ref{eq:hyperbolic sum}) in our limit:
\begin{equation}
\label{eq:sum hyperbolic in case of infinite kx'}
\begin{gathered}
    S_{\text{hyp}}(k_x^{'}, N_x) = \frac{1}{4} \left(\frac{\exp k_x^{'} (2 N_x + 1)}{\exp k_x^{'}} - (2 N_x + 1) \right) \\
    = \lim_{\eta \rightarrow 0} \frac{1}{4} \left(\frac{(2 \eta)^{-(2 N_x + 1)}}{(2 \eta)^{-1}} - (2 N_x + 1) \right) \\
    = \lim_{\eta \rightarrow 0} \frac{1}{4} \left((2 \eta)^{-2 N_x} - (2 N_x + 1) \right) = \frac{(2 \eta)^{-2 N_x}}{4}.
\end{gathered}
\end{equation}
The wave functions (first of Eq. (\ref{eq:zGNR localized wave function final})) now have the form
\begin{equation}
\label{eq:zGNR special case}
\begin{gathered}
    \mqty(\psi_{A,\textbf{i}} \\ \psi_{B,\textbf{i}} ) = \lim_{\eta \rightarrow 0} \frac{(-1)^n e^{i \left( \frac{\pi}{2} - \eta \right) m}}{\sqrt{N_y \cdot (2 \eta)^{-2 N_x}/4}} \mqty(s \cdot \frac{1}{2} e^{\left( - \ln 2 \eta \right)  n}     \\      \frac{1}{2} e^{\left( - \ln 2 \eta \right)  (N_x-n)} ) \\
    = \lim_{\eta \rightarrow 0} \frac{(-1)^{n} e^{i \frac{\pi}{2} m}}{\sqrt{N_y}} \mqty(s \cdot (2 \eta)^ {N_x - n}     \\    (2 \eta)^{n} ).
\end{gathered}
\end{equation}
From the last expression, we see that for the limit $\eta \rightarrow 0$ we have nonzero wave functions $\psi_{A,\textbf{i}}$ only for $n = N_x$ (on the right zigzag edge) and nonzero wave functions $\psi_{B,\textbf{i}}$ only for $n = 0$ (on the left one).  So these wave functions in the limit can be written as entirely localized states:
\begin{equation}
\label{eq:zGNR special case final}
    \mqty(\psi_{A,\textbf{i}} \\ \psi_{B,\textbf{i}} ) = \frac{(-1)^n \cdot i^m }{\sqrt{N_y}} \mqty( s \cdot \delta_{N_x,n} \\ \delta_{0,n}),
\end{equation}
where $\delta_{i,j}$ is the Kronecker delta.

It is important to note that we find to real wave functions in the case of even $N_x$.
In the case of odd $N_x$, at the left edge wave functions are real; the right edge wave functions are imaginary.
This can be changed by renormalization.

\begin{figure}
    \centering
    \includegraphics[width=0.9\linewidth]{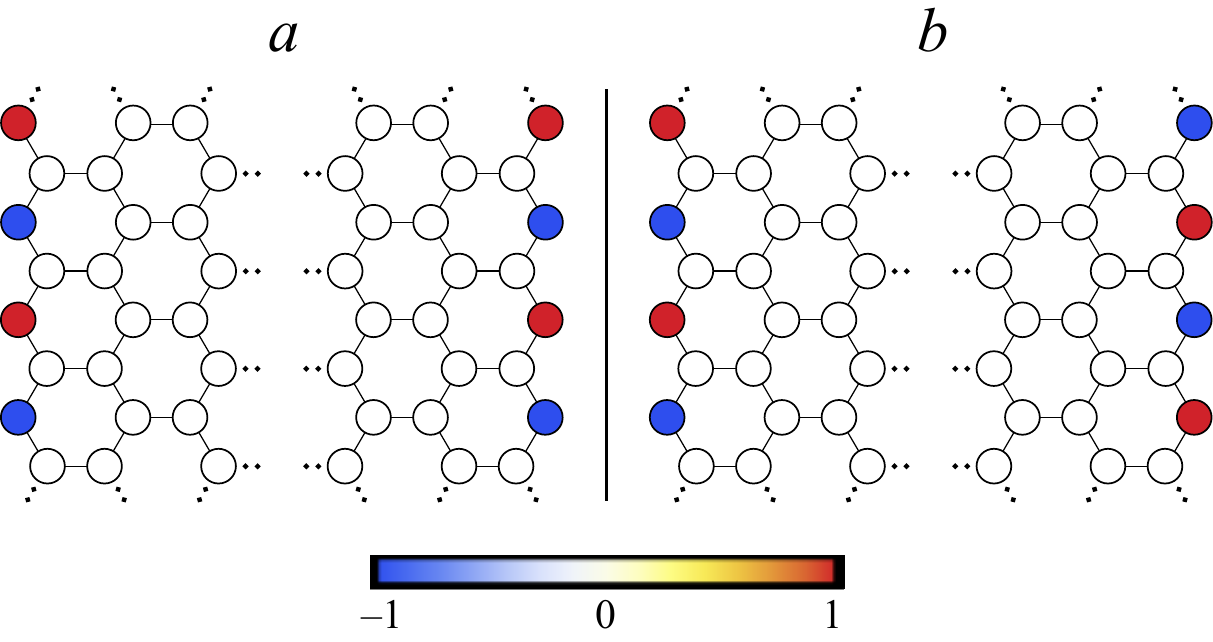}
    \caption{Wave function of zigzag nanoribbon in entirely localized state (\ref{eq:zGNR special case final}) for even $N_x$: (a) describes conduction band, (b) describes valence band. The amplitudes are normalized by the maximum value of the respective wave functions.}
    \label{fig: zigzag nanoribbon wave function}
\end{figure}

There are two basic entirely localized edge states depending on the parameter $s$ which describe the valence (conduction) band.
They are illustrated in Fig. \ref{fig: zigzag nanoribbon wave function}.
However, because of their identical energies, one can construct a superposition of states and obtain state entirely localized on only one zigzag edge (left or right one).

Note that result (\ref{eq:zGNR special case final}) does not depend on $N_x$.
This means that these states have identical form for all allowed values of $N_x$ (even) and in the limit of the semi-infinite plane, this result coincides with \cite{fujita1996peculiar}.

When we look at the right-hand limit $k_y = \frac{\pi}{2} + \eta$, where $\eta > 0$ (see Appendix \ref{app:ky = pi/2}), we obtain a zero-energy state with wave functions
\begin{equation}
\label{eq:zGNR special case final right limit}
    \mqty(\psi_{A,\textbf{i}} \\ \psi_{B,\textbf{i}} ) = \frac{i^m}{\sqrt{N_y}} \mqty( s \cdot \delta_{N_x,n} \\ \delta_{0,n}).
\end{equation}
This result differs by a factor of $(-1)^n$ compared to the left-hand limit (\ref{eq:zGNR special case final}).
In the case of even $N_x$, the resulting wave functions coincide.
For odd $N_x$ they also coincide if one compares nonzero valence band wave functions (\ref{eq:zGNR special case final}) to nonzero conduction band wave functions (\ref{eq:zGNR special case final right limit}) and vice versa.
However, conduction and valence bands at the point $k_y = \pi / 2$ are experimentally indistinguishable because they have identical (zero) energy.
Hence, wave functions in both forms (\ref{eq:zGNR special case final}) and (\ref{eq:zGNR special case final right limit}) are the same.

Note that this entirely localized state exists in the case of the carbon nanotube with open zigzag edges for $N_y$ divisible by 4. This can be seen from the allowed values of wavenumber $k_y$ (\ref{eq:zGNR ky}).

\subsubsection{Cases $k_y = k_y^c, \pi - k_y^c$}

These values of wavenumber $k_y$ are on the borders between the case of all real roots and the case when one root of Eq. (\ref{eq:numerical equation}) is complex.
All $N_x - 1$ real roots of Eq. (\ref{eq:numerical equation}) can be obtained numerically and the corresponding wave functions can be presented in the form (\ref{eq:zGNR extended wave function final}).
The main interest is to describe wave functions and energies for the last root $k_x$, which tends to zero for $k_y \rightarrow \pi - k_y^c$ and tends to $\pi$ for $k_y$ approaching $k_y^c$.
These points can be called transition points because here bulk states transform to edge ones and vice versa.

First, let us look at the left transition point $k_y = k_y^c$.
We start approaching this point from the left ($k_y = k_y^c - \eta$, $\eta \rightarrow 0^+$).
The root of Eq. (\ref{eq:numerical equation}) that we are interested in tends to $\pi$ also from the left ($k_x = \pi - \lambda$, $\lambda > 0$).

We need to simplify the summation function (\ref{eq:sum sin^2}). It is possible to do by expanding into series formula (\ref{eq:sum sin^2}) or substitute $k_x = \pi - \lambda$ at summation. We use the second approach:
\begin{equation}
\label{eq: sum sin^2 transition point}
\begin{gathered}
    S(k_x, N_x) = \lim_{\lambda \rightarrow 0} \sum_{n=1}^{N_x} \sin^2 (\pi - \lambda) n
    = \sum_{n=1}^{N_x} (-1)^{2n} (\lambda n)^2 \\
    = \lambda^2 \sum_{n=1}^{N_x} n^2 = \lambda^2 \frac{N_x(N_x+1)(2N_x + 1)}{6}.
\end{gathered}
\end{equation}

Then we need to simplify the sign function $s_1$ (\ref{eq:s1}):
\begin{equation}
\begin{gathered}
    s_1 = \lim_{\lambda \rightarrow 0} \text{sign} \left(\sin (\pi - \lambda) (N_x+1) \right) \\
    = \lim_{\lambda \rightarrow 0} \text{sign} \left(- (-1)^{N_x + 1}\sin \lambda (N_x+1) \right) = (-1)^{N_x}.
\end{gathered}
\end{equation}

The next step is to calculate the function $\epsilon (k_x, k_y)$ (\ref{eq:special case energies}) which is proportional to the energy $E_s$:
\begin{equation}
\label{eq:k_y^c energy}
\begin{gathered}
   \epsilon (\pi, k_y^c) = \medmath{\lim_{\lambda,\eta \rightarrow 0} \sqrt{3 + 4 \cos{(\pi - \lambda)} \cos{(k_y^c - \eta)} + 2 \cos{2 (k_y^c - \eta)}}} \\
    = \sqrt{3 - 4 \cos{k_y^c} + 2 \cos{2 k_y^c}} = \frac{1}{N_x + 1}.
\end{gathered}
\end{equation}
One can see that the energy $E_s = s \cdot t \cdot \epsilon (\pi, k_y^c)$ is different from zero, but approaches it in the case of the infinite (semi-infinite) system.

The last step is to simplify the wave functions (\ref{eq:zGNR extended wave function final}):
\begin{equation}
\label{eq:zGNR k_y^c case}
\begin{gathered}
    \mqty(\psi_{A,\textbf{i}} \\ \psi_{B,\textbf{i}} ) = \lim_{\lambda,\eta \rightarrow 0} \frac{e^{i (k_y^c - \eta) m}}{\sqrt{N_y} \lambda \sqrt{N_x(N_x+1)(2N_x + 1)/6}} \\
    \cdot \mqty((-1)^{N_x} s \sin \left( (\pi - \lambda) n \right)  \\
    \sin \left( (\pi - \lambda) (N_x-n) \right) ) \\
    = \frac{\sqrt{6} e^{i k_y^c m}}{\lambda \sqrt{N_y N_x(N_x+1)(2N_x + 1)}} \mqty(- s (-1)^{N_x + n} \lambda n  \\
    - (-1)^{N_x - n} \lambda (N_x-n)) \\
    = \frac{(-1)^{N_x+ n + 1} \sqrt{6} e^{i k_y^c m}}{\sqrt{N_y N_x(N_x+1)(2N_x + 1)}} \mqty(s \cdot n  \\
    N_x-n).
\end{gathered}
\end{equation}
Here, at the last step we used the identity $(-1)^{- 2n} = 1$.
These wave functions can be renormalized to remove a factor $(-1)^{N_x+1}$.
They also coincide with wave functions obtained from the right limit at the point after renormalization (for more details see Appendix \ref{app:Right limit of wave functions at k_y^c}).

Another border point $k_y = \pi - k_y^c$ can be treated by analogy (see Appendices \ref{app:Left limit of wave functions at pi - k_y^c} and \ref{app:Right limit of wave functions at pi - k_y^c} for the left and the right limits respectively). As a result, we have energies $E_s$ identical to the ones in the case $k_y = k_y^c$, and wave functions in the form
\begin{equation}
\label{eq:zGNR pi-k_y^c wave function}
    \mqty(\psi_{A,\textbf{i}} \\ \psi_{B,\textbf{i}} ) = \frac{(-1)^m \sqrt{6} e^{-i k_y^c m}}{\sqrt{N_y N_x(N_x+1)(2N_x + 1)}} \mqty(s \cdot n  \\ N_x-n).    
\end{equation}

Comparing the resulting wave functions for two transition points $k_y = k_y^c$ (\ref{eq:zGNR k_y^c case}) and $k_y = \pi - k_y^c$ (\ref{eq:zGNR pi-k_y^c wave function}) one can note their linear dependence on the horizontal index $n$, but in the first case there is an extra sign function that distinguish between odd and even horizontal cells.
In the second, there is a sign function which changes for different vertical positions of the cell $m$.
Now we try to show that these sign functions work coherently: if one has value $+1$, the other one also equals $+1$ and vice versa for our choice of cell numeration.
If one looks at Fig. \ref{fig: zigzag nanoribbon} with our numeration convention it is possible to note that for any cell, $n$ and $m$ are both odd or are both even.
For shifted numeration one can get the opposite result: when $n$ is even $m$ is odd and vice versa.
For this case, the minus can be absorbed by renormalization.
We could find only one work \cite{Saroka2017Optics} where this transition point was considered, and the obtained result coincides with (\ref{eq:zGNR pi-k_y^c wave function}).

Note that obtained wave functions can be applied both for nanoribbons and nanotubes. The only difference is that for infinite nanoribbons longitudinal wavenumber is continuous, and it is discrete for nanotubes.

\section{Finite sample} \label{sec:Finite sample}

\begin{figure}
    \centering
    \includegraphics[width=0.99\linewidth]{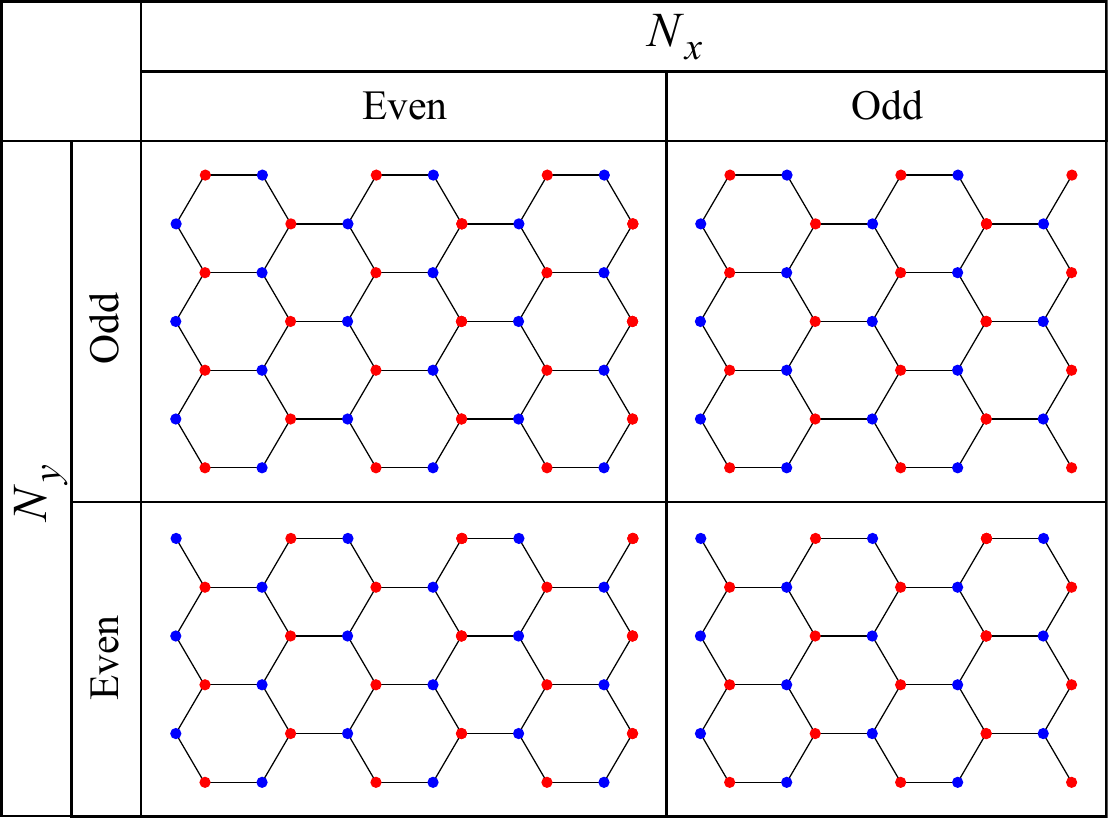}
    \caption{Possible rectangular geometries of finite size honeycomb lattice.}
    \label{fig: finite system}
\end{figure}

The easiest way to derive wave functions for a finite sample of rectangular geometry (Fig. \ref{fig: finite system}) is to use the results of previous sections as a base.
For the case of an armchair nanoribbon we found only one possible type of state (extended ones), but for zigzag nanoribbons we found a big variety of possible states.
Therefore, we will use wave functions for zigzag nanoribbons which describe running waves in the $+y$ direction and superpose them with wave functions for zigzag nanoribbons which describe running waves in the $-y$ direction (they are obtained by changing $k_y$ to $-k_y$ in the formulas of Sec. \ref{sec:wave functions for zigzag nanoribbons}).
The resulting wave functions automatically satisfy the boundary conditions for zigzag nanoribbons (\ref{eq: BC for zigzag}) and after imposing boundary conditions for armchair nanoribbons (\ref{eq: BC for armchair}), terms $e^{i k_y m}$ transform to $\sin{k_y m}$ like in Sec. \ref{sec:wave functions for armchair nanoribbons}.
Normalization constants should be also changed: instead of factors $1/\sqrt{N_y}$ there will be $1/\sqrt{S(k_y,N_y)}$.

The eigenvalue problem (\ref{eq:Schrodinger}) for a finite sample with $N_x \cross N_y$ sites should have $N_x \cross N_y$ eigenvalues and eigenvectors.
If we take all possible values $k_y$ (\ref{eq:ky}) from the armchair nanoribbon solution and solve the transcendental equation (\ref{eq:numerical equation}) for each of them, finally we get $2 N_x N_y$ solutions.
Here the doubling comes from valence and conduction band solutions.
The solution to this problem is as follows: pairs of wavenumbers $(k_x,k_y)$ and $(\pi - k_x, \pi - k_y)$ describe identical states (see Appendix \ref{app:finite sample proof of change} for the proof).
Therefore, we can halve the number of such pairs, considering that \cite{yorikawa2021edge}
\begin{equation}
\label{eq:ky for finite system}
    k_y = \frac{\pi j_y}{N_y + 1}, \quad j_y = 1, 2, 3, \ldots \frac{N_y + 1}{2} \quad (0 < k_y \leq \frac{\pi}{2})
\end{equation}
and
\begin{align}
\label{eq:kx range for finite system}
    & 0 < k_x < \pi                \qquad (0 < k_y < \frac{\pi}{2}), \\
    \label{eq:kx constraints}
    & 0 < k_x \leq \frac{\pi}{2}   \qquad (k_y = \frac{\pi}{2}),
\end{align}
which are to be found from Eq. (\ref{eq:numerical equation}).

Note that the entirely localized states which were found in zigzag nanoribbons for $k_y = \pi / 2$ in Sec. \ref{sec:Case k_y = pi/2} also exist in finite honeycomb systems.
However, the special value of wavenumber $k_y = \pi / 2$ can't be obtained for systems with even $N_y$ (see (\ref{eq:ky for finite system}) and forbidden geometries in the second line of Fig. \ref{fig: finite system}).

We divide further explanations into subsections where we find wave functions for different regions (points) of wavenumber $k_y$.

\subsubsection{Case $k_y \in \left( 0 ; k_y^c \right)$}

All solutions in this region of $k_y$ correspond to extended states. Energies can be calculated by formulas (\ref{eq:energies}). Wave functions for zigzag nanoribbon (\ref{eq:zGNR extended wave function final}) are modified to
\begin{equation}
\label{eq:extended states finite system}
    \mqty(\psi_{A,\textbf{i}} \\ \psi_{B,\textbf{i}} ) = \frac{\sin{k_y m}}{\sqrt{S(k_x, N_x) S(k_y, N_y)}} \mqty(s_1 s \sin (k_x n)  \\
    \sin (k_x (N_x-n))),
\end{equation}
where the sign function $s_1$ is still defined by (\ref{eq:s1}). The possible $N_x$ real values of wavenumber $k_x$ for each $k_y$ can be found numerically from (\ref{eq:numerical equation}).

This result is identical to the wave functions obtained in paper \cite{yorikawa2021edge}.
If one wants to compare this result to wave functions obtained for unit cell consisting of four atoms \cite{malysheva2008spectrum, ruseckas2011spectrum} they need to know that in the case Brillouin zone is smaller.
This leads to the appearance of additional dispersion branches which are artificially made by introducing a factor $\pm 1$ in the second term of $\epsilon (k_x, k_y)$ (\ref{eq:energies}) (the second branch appears for $k_x \rightarrow \pi + k_x$ in our notation).
This is the reason why direct comparison of our result to the results in \cite{malysheva2008spectrum, ruseckas2011spectrum} is difficult.
We note that their wave functions coincide with our result (\ref{eq:extended states finite system}) if we do not take into account the factor which describes additional branches.

There are $N_x -1$ real roots of the transcendental Eq. (\ref{eq:numerical equation}) with corresponding extended states described by wave functions (\ref{eq:extended states finite system}) for the region of $k_y \in \left[ k_y^c; \pi/2 \right)$.
Next, we discuss the only complex root of Eq. (\ref{eq:numerical equation}) and the corresponding wave functions in the above-mentioned region for wavenumber $k_y$.
We will talk later about the point $k_y = \pi / 2$ and discuss both real and complex roots.

\subsubsection{Case $k_y \in \left( k_y^c ; \pi/2 \right)$}
We obtain these wave functions by modifying wave functions for zigzag nanoribbons (\ref{eq:zGNR localized wave function final}) for the region $k_y \in \left( k_y^c ; \pi/2 \right)$:
\begin{equation}
\label{eq:localized wave function finite system}
    \mqty(\psi_{A,\textbf{i}} \\ \psi_{B,\textbf{i}} ) = \frac{(-1)^n \sin{k_y m}}{\sqrt{S_{\text{hyp}}(k_x^{'}, N_x) S(k_y, N_y)}} \mqty(s \sinh (k_x^ {'}  n)     \\      \sinh (k_x^ {'}  (N_x-n)) ),
\end{equation}
where $k_x^ {'}$ is a positive root of the following equation:
\begin{equation}
    \sinh{k_x^ {'} N_x} - 2 \cos{k_y} \sinh{\left(k_x^ {'} (N_x + 1) \right)} = 0.
\end{equation}
The same result can be derived directly from wave functions (\ref{eq:extended states finite system}) with the complex plane continuation $k_x = \pi + i k_x^{'}$ by analogy with the calculation in Appendix \ref{app:Derivation of wave functions for localized case}.
Corresponding energies have the form
\begin{equation}
\label{eq:energies for finite system in localized state}
    E_s = s \cdot t \sqrt{3 - 4 \cosh{k_x^ {'}} \cos{k_y} + 2 \cos{2 k_y}}.
\end{equation}

The result (\ref{eq:localized wave function finite system}) agrees with the recent work \cite{yorikawa2021edge} that explored localized states in finite systems with rectangular geometry.

\subsubsection{Case $k_y = k_y^c$}
In this case $k_x$ tends to $\pi$. We can use wave functions of (\ref{eq:zGNR k_y^c case}) or (\ref{eq:zGNR pi-k_y^c wave function}) as a basis, where we need to change $e^{i k_y^c m}$ (or $e^{-i k_y^c m}$) to $\sin{k_y^c m}$ together with changing the normalization coefficient. We rewrite (\ref{eq:zGNR k_y^c case}):
\begin{equation}
\label{eq:k_y^c wave function finite sample}
    \mqty(\psi_{A,\textbf{i}} \\ \psi_{B,\textbf{i}} ) = \frac{(-1)^{N_x+ n + 1} \sqrt{6} \sin{k_y^c m}}{\sqrt{N_x(N_x+1)(2N_x + 1) S(k_y^c, N_y)}} \mqty(s \cdot n  \\ N_x-n).    
\end{equation}
One can obtain this result as a left-hand limit to the point $k_y = k_y^c$ for finite sample extended wave functions (\ref{eq:extended states finite system}) or a right-hand limit of localized wave functions (\ref{eq:localized wave function finite system}).
If one wants to check coincidence of wave functions after replacement $(k_x,k_y) \rightarrow (\pi - k_x, \pi - k_y)$, straightforward approach here can't be done.
In this case, we need to look at the right-hand limit of wave functions (\ref{eq:extended states finite system}) at the point $k_y = \pi - k_y^c$ which is treated as in Appendix \ref{app:Right limit of wave functions at pi - k_y^c}, and we come to the result
\begin{equation}
\label{eq:k_y^c wave function finite sample 2}
    \mqty(\psi_{A,\textbf{i}} \\ \psi_{B,\textbf{i}} ) = \frac{(-1)^{m+1} \sqrt{6} \sin{k_y^c m}}{\sqrt{N_x(N_x+1)(2N_x + 1) S(k_y^c, N_y)}} \mqty(s \cdot n  \\ N_x-n).    
\end{equation}
The explanation for why factor $(-1)^m$ in (\ref{eq:k_y^c wave function finite sample}) is the same as $(-1)^n$ in (\ref{eq:k_y^c wave function finite sample 2}) was shown at the end of Sec. \ref{sec:wave functions for zigzag nanoribbons}.

Corresponding energies coincide with energies found for zigzag nanoribbons in the case $k_y = k_y^c$:
\begin{equation}
    E_s = \frac{s \cdot t}{N_x + 1}.
\end{equation}

\subsubsection{Case $k_y = \pi/2$}
This case is only possible for odd $N_y$ (as $k_y$ values have form (\ref{eq:ky for finite system})).
We start by describing a sample with odd $N_x$ and $N_y$.
Possible $k_x$ values can be found analytically from (\ref{eq:sin kxNx=0}) by taking into account constraints (\ref{eq:kx constraints}):
\begin{equation}
    k_x = \frac{\pi}{N_x} j_x, \quad j_x = 1, 2, 3, \ldots \frac{N_x - 1}{2}.
\end{equation}
The corresponding extended wave functions are described by formulas (\ref{eq:extended states finite system}).

\begin{figure}
    \centering
    \includegraphics[width=0.4\linewidth]{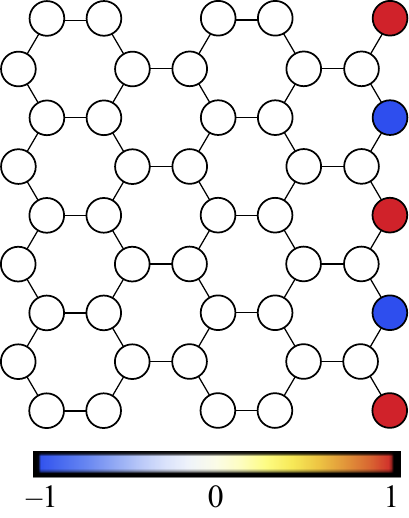}
    \caption{Wave function for the entirely localized state in finite honeycomb system (\ref{eq:finite system zero energy wave function}). The amplitudes are normalized by the maximum value of the respective wave functions.}
    \label{fig: finite wave function}
\end{figure}

Now let us count the number of roots of the eigenvalue problem (\ref{eq:Schrodinger}) for a rectangular system with $N_x N_y$ sites. For $k_y \in \left( 0 ; \frac{\pi}{2} \right)$ there are $N_x (N_y - \frac{1}{2})$ solutions, for $k_y = \frac{\pi}{2}$ and $k_x \in \left( 0 ; \frac{\pi}{2} \right)$ we have $N_x - 1$ solutions (doubling comes from the two possible bands).
So there is only one root remaining that we need to find.
We expect to find it as a limit of a complex one like we found for the entirely localized state in Sec. \ref{sec:wave functions for zigzag nanoribbons} for $k_y = \pi / 2$.
Considering the left-hand limit $k_y = \frac{\pi}{2} - \eta$ (where $\eta \rightarrow 0^+$) one can obtain by analogy with Sec. \ref{sec:wave functions for zigzag nanoribbons} $k_x = \pi \pm i k_x^ {'}$ ($k_x^ {'}$ tends to infinity). This state has zero energy (see (\ref{eq:zero energy})) and the corresponding wave function can be written using the Kronecker delta function:
\begin{equation}
\label{eq:finite system zero energy wave function notfinal}
    \mqty(\psi_{A,\textbf{i}} \\ \psi_{B,\textbf{i}} ) = \frac{(-1)^n \cdot \sin{\frac{\pi}{2} m} }{\sqrt{N_y}} \mqty( s \cdot \delta_{N_x,n} \\ \delta_{0,n}).
\end{equation}
However, this result can be significantly simplified.
When one looks at our numeration convention (Fig. \ref{fig: zigzag nanoribbon}) and geometry of the system we study now (odd $N_x$ and $N_y$, upper right corner of Fig. \ref{fig: finite system}) they can note that the wave function at $B$ sites is zero.
It comes from the factor $\delta_{0,n} \sin{\frac{\pi}{2} m}$.
The first function is nonzero only for $n = 0$ (left edge), but at the left edge $m$ is even and $\sin{\frac{\pi}{2} m} = 0$.
Finally the wave function can be written as
\begin{equation}
\label{eq:finite system zero energy wave function}
    \mqty(\psi_{A,\textbf{i}} \\ \psi_{B,\textbf{i}} ) = \frac{\delta_{N_x,n} \sin{\frac{\pi}{2} m} }{\sqrt{(N_y + 1)/2}} \mqty( 1 \\ 0),
\end{equation}
where we have removed the factor $(-1)^n$ because it was multiplied by a function which is nonzero at the right edge ($n = N_x = \text{const}$) and changed the normalization constant, because the previous one was derived assuming that the wave function is localized on both left and right edges, but for this geometry it can be localized only on the right edge.
Here the band index $s$ is also absent because it is easy to show that after renormalization wave functions both conductance and valence bands are identical.
Therefore, we have found the last solution of the eigenvalue problem (\ref{eq:Schrodinger}) in the case of odd $N_x$.
This wave function is illustrated in Fig. \ref{fig: finite wave function}.

Such rectangular geometry (odd $N_x$ and $N_y$) was studied numerically \cite{yorikawa2021edge} and this entirely localized state was also observed.

We now switch to the case of even $N_x$ and odd $N_y$. Possible $k_x$ values are as follows:
\begin{equation}
    k_x = \frac{\pi}{N_x} j_x, \quad j_x = 1, 2, 3, \ldots \frac{N_x}{2}.
\end{equation}
The corresponding wave functions (\ref{eq:extended states finite system}) describe extended states.

If one counts the number of roots like we did earlier, they will find that we already have the correct amount of solutions for the problem (\ref{eq:Schrodinger}).
Consequently, there are no entirely localized states for rectangular geometry with even $N_x$ and odd $N_y$ (upper left corner of Fig. \ref{fig: finite system}).
Let us also show it based on the result (\ref{eq:finite system zero energy wave function notfinal}): one can see that both Kronecker delta functions are nonzero for either $n = 0$ or $n = N_x$ (which is even), on both these edges $m$ is even and $\sin{\frac{\pi}{2} m} = 0$ (because of our indexation convention (Fig. \ref{fig: zigzag nanoribbon})), and we have zero wave function everywhere. This is nonphysical and we can say that a zero energy state for this geometry is not realizable.

All obtained solutions are numerically verified to satisfy  Schr$\ddot{\text{o}}$dinger equation (\ref{eq:Schrodinger}) and both normalization conditions (\ref{eq:normalization}).

\section{Discussion}
Electronic properties of boundaries of the honeycomb lattice are one of the most basic problems of systems with boundaries, physically realized in graphene, artificial honeycomb structures and ultracold atoms in honeycomb optical lattices.
The most studied example of a boundary is
the semi-infinite graphene sheet with zigzag edge \cite{fujita1996peculiar, nakada1996edge, wakabayashi1999electronic, deng2014edge}.
It was shown that this has a band of zero-energy surface states for $k_y \in \left( \frac{1}{3} \pi; \frac{2}{3} \pi \right)$ in our notation, with dimensionless penetration length given by $\lambda = -1 / \ln{\lvert 2 \cos{k_y} \rvert}$.
For this system, it was also shown that there exists a state with an entirely localized wave function at the edge for $k_y = \pi / 2$ (similar to the states we found for zigzag nanoribbons and nanotubes in Fig. \ref{fig: zigzag nanoribbon wave function}, rectangular graphene nanoflake in Fig. \ref{fig: finite wave function}).
For zigzag nanoribbons, these authors discussed edge states with $E \approx 0$ \cite{nakada1996edge}, but a state with zero energy was not mentioned.

A zero energy state was numerically studied for finite graphene nanoflakes \cite{yorikawa2021edge} and the state illustrated in Fig. \ref{fig: finite wave function} was found.
The reason for its appearance is thought to be sublattice imbalance (number of sites of $A$ and $B$ sublattices differs by 1).
It was also noted that this state can not be described by the usual extended or localized wave functions representations.
However, we showed that entirely localized states should be treated as a limit of localized states when $k_y \rightarrow \pi / 2$ and $E \rightarrow 0$.

In this paper, exact electron spectrum and wave functions based on the tight-binding model for armchair and zigzag nanoribbons and nanotubes, rectangular graphene nanoflakes have been presented.
We showed that localized states can exist in zigzag nanoribbons, zigzag nanotubes and rectangular graphene nanoflakes.
Entirely localized states with zero energy (when wave function is nonzero only at the edge sites) were found in zigzag nanoribbons, zigzag nanotubes with number of sites along zigzag edge divisible by 4, and rectangular graphene nanoflakes with odd number of sites along zigzag and armchair edges.
We described the transition point between extended and localized states.
Here wave functions can be written as linear functions of the horizontal index $n$ times the sign changing function $(-1)^n$.
It looks like a localized state with critical (i.e. infinite) penetration length.

After this work was completed we became aware of the recent studies of graphene ribbons \cite{kasturirangan2022disordered}.

\begin{acknowledgments}
We thank Chris Halcrow, Albert Samoilenka and Mats Barkman for useful discussions. 
This research was financially supported by
the Knut and Alice Wallenberg Foundation through the Wallenberg Center for Quantum Technology (WACQT)
and Swedish Research Council Grants 2016-06122, 2018-03659.
\end{acknowledgments}

\bibliography{references.bib}

\clearpage

\appendix

\section{Derivation of wave functions for localized case} \label{app:Derivation of wave functions for localized case}
First, let us derive wave functions for the case of complex plane continuation in the form $k_x = \pi + i k_x^{'}$, where $k_x^{'} > 0$. We start from simplifying the sign function $s_1$ (\ref{eq:s1}) using the identity $\sin{(\pi + i k_x^{'}) n} = i (-1)^n \sinh{k_x^{'} n}$:
\begin{equation}
\begin{gathered}
    s_1 = \frac{ \sqrt{\sin^2 (\pi + i k_x^{'})} }{\sin (\pi + i k_x^{'})} \cdot \frac{ \sin ((\pi + i k_x^{'}) (N_x+1))}{ \sqrt{\sin^2 ((\pi + i k_x^{'}) (N_x+1)) }} \\
    = \frac{\sqrt{(-i \sinh{k_x^{'}})^2}}{-i \sinh{k_x^{'}}} \cdot \frac{i (-1)^{N_x + 1} \sinh{k_x^{'} (N_x + 1)}}{\sqrt{(i (-1)^{N_x + 1} \sinh{k_x^{'} (N_x + 1)})^2}} \\
    = (-1)^{N_x}.
\end{gathered}
\end{equation}
Now we can rewrite the wave functions from Eq. (\ref{eq:zGNR extended wave function}):
\begin{equation}
\label{eq:app wave function extended k_x = pi + i k_x'}
\begin{gathered}
    \mqty(\psi_{A,\textbf{i}} \\ \psi_{B,\textbf{i}} ) = - \sqrt{2} i c_3^{'} e^{i ((\pi + i k_x^{'}) N_x + k_y m)} \\
    \cdot \mqty((-1)^{N_x} s \sin ((\pi + i k_x^{'}) n)  \\    \sin ((\pi + i k_x^{'}) (N_x-n))) \\
    = (-1)^{N_x + 1} \sqrt{2} i c_3^{'} e^{- k_x^{'} N_x + i k_y m} \\
    \cdot \mqty((-1)^{N_x} s \cdot i (-1)^n \sinh{k_x^{'} n}  \\    i (-1)^{N_x - n} \sinh{k_x^{'} (N_x - n)}) \\
    = (-1)^{n + 2} \sqrt{2} c_3^{'} e^{- k_x^{'} N_x + i k_y m} \\
    \cdot \mqty(s (-1)^{2 N_x} \sinh{k_x^{'} n}  \\    (-1)^{2N_x - 2n} \sinh{k_x^{'} (N_x - n)}) \\
    = (-1)^{n} \sqrt{2} c_3^{'} e^{- k_x^{'} N_x + i k_y m} \mqty(s \sinh{k_x^{'} n}  \\   \sinh{k_x^{'} (N_x - n)}).
\end{gathered}
\end{equation}

In the second possible case of continuation ($k_x = 0 + i k_x^{'}$, where $k_x^{'} > 0$) we apply the identity $\sin{i k_x^{'} n} = i \sinh{k_x^{'} n}$. First we rewrite $s_1$ as
\begin{equation}
\begin{gathered}
    s_1 = \frac{ \sqrt{\sin^2 (i k_x^{'})} }{\sin (i k_x^{'})} \cdot \frac{ \sin (i k_x^{'} (N_x+1))}{ \sqrt{\sin^2 (i k_x^{'} (N_x+1)) }} \\
    = \frac{\sqrt{(i \sinh{k_x^{'}})^2}}{i \sinh{k_x^{'}}} \cdot \frac{i \sinh{k_x^{'} (N_x + 1)}}{\sqrt{(i \sinh{k_x^{'} (N_x + 1)})^2}} = 1.
\end{gathered}
\end{equation}
The next step is to  present wave functions (\ref{eq:zGNR extended wave function}) in the following form:
\begin{equation}
\begin{gathered}
    \mqty(\psi_{A,\textbf{i}} \\ \psi_{B,\textbf{i}} ) = - \sqrt{2} i c_3^{'} e^{i (i k_x^{'} N_x + k_y m)} \mqty(1 \cdot s \sin (i k_x^{'} n)  \\    \sin (i k_x^{'} (N_x-n))) \\
    = - \sqrt{2} i c_3^{'} e^{- k_x^{'} N_x + i k_y m} \mqty(s \cdot i \sinh{k_x^{'} n}  \\    i \sinh{k_x^{'} (N_x - n)}) \\
    = \sqrt{2} c_3^{'} e^{- k_x^{'} N_x + i k_y m} \mqty(s \sinh{k_x^{'} n}  \\   \sinh{k_x^{'} (N_x - n)}).
\end{gathered}
\end{equation}

Here we used not normalized expression for extended wave functions Eq. (\ref{eq:zGNR extended wave function}) instead of normalized one (\ref{eq:zGNR extended wave function final}) in order not to transform summation formula (\ref{eq:sum sin^2}), because anyway we need to renormalize the resulting wave functions.

\section{Right limit of wave functions and energies for zigzag nanoribbons in the case $k_y = \pi / 2$} \label{app:ky = pi/2}
For this type of limit we present wavenumber $k_y$ as $\frac{\pi}{2} + \eta$, where $\eta \rightarrow 0^+$.
Acting by analogy with Section \ref{sec:Case k_y = pi/2} we can write that wavenumber $k_x$ has the form $0 \pm i k_x^{'}$, where $k_x^{'}$ tends to infinity. The eigenvalue equation (second of Eq.(\ref{eq:hyperbolic equations})) in these limits can be written as
\begin{equation}
    \frac{1}{2} e^{k_x^ {'} N_x} + 2 \cos{\left( \frac{\pi}{2} + \eta \right)} \cdot \frac{1}{2} e^{k_x^ {'} (N_x + 1)} = 0,
\end{equation}
with solution
\begin{equation}
    k_x^ {'} = - \ln 2 \eta.
\end{equation}
This dependence $k_x^ {'}$ on $\eta$ is identical in the left-hand limit.

Eigenenergies $E_s$ tend to zero, because the function $\epsilon (k_x^ {'}, \frac{\pi}{2})$ (\ref{eq:special case energies}) tends to zero:
\begin{equation}
\begin{gathered}
   \epsilon (k_x^ {'}, \frac{\pi}{2}) = \sqrt{3 + 4 \cdot \frac{1}{2} e^{k_x^ {'}} \cos{k_y} + 2 \cos{2 k_y}} \\
    = \lim_{\eta \rightarrow 0} \sqrt{3 + 2 e^{-\ln{2 \eta}} \cos{(\frac{\pi}{2} + \eta)} + 2 \cos{2 (\frac{\pi}{2} + \eta)}} \\
    = \lim_{\eta \rightarrow 0} \sqrt{3 - 2 \cdot (2 \eta)^{-1} \eta + 2 \cdot (-1)} = 0.
\end{gathered}
\end{equation}

The summation function (\ref{eq:hyperbolic sum}) inside wave functions (\ref{eq:zGNR localized wave function final}) simplifies. It has a form identical to (\ref{eq:sum hyperbolic in case of infinite kx'}). Finally, the wave functions (second of Eq. (\ref{eq:zGNR localized wave function final})) become
\begin{equation}
\label{eq:app zGNR special case}
\begin{gathered}
    \mqty(\psi_{A,\textbf{i}} \\ \psi_{B,\textbf{i}} ) = \lim_{\eta \rightarrow 0} \frac{e^{i \left( \frac{\pi}{2} + \eta \right) m}}{\sqrt{N_y \cdot (2 \eta)^{-2 N_x}/4}} \mqty(s \cdot \frac{1}{2} e^{\left( - \ln 2 \eta \right)  n}     \\      \frac{1}{2} e^{\left( - \ln 2 \eta \right)  (N_x-n)} ) \\
    = \lim_{\eta \rightarrow 0} \frac{e^{i \frac{\pi}{2} m}}{\sqrt{N_y}} \mqty(s \cdot (2 \eta)^ {N_x - n}     \\    (2 \eta)^{n} ).
\end{gathered}
\end{equation}
Similarly to Eq. (\ref{eq:zGNR special case}) and (\ref{eq:zGNR special case final}) we can rewrite the result (\ref{eq:app zGNR special case}) using Kronecker delta functions:
\begin{equation}
    \mqty(\psi_{A,\textbf{i}} \\ \psi_{B,\textbf{i}} ) = \frac{i^{m}}{\sqrt{N_y}} \mqty( s \cdot \delta_{N_x,n} \\ \delta_{0,n}).
\end{equation}

\section{Simplification of wave functions for zigzag nanoribbons in cases $k_y = k_y^c, \pi - k_y^c$} \label{app:k_y^c}
Here we provide the rest of the limits of wave functions at transition points $k_y = k_y^c, \pi - k_y^c$ that are not mentioned in the main text.

\subsection{Right limit of wave functions at the point $k_y = k_y^c$} \label{app:Right limit of wave functions at k_y^c}
We are interested in results for complex plane continuation of type $k_x = \pi \pm i k_x^{'}$. We represent wavenumber $k_y$ in the form $k_y^c + \eta$, where $\eta \rightarrow 0^+$. In this limit, $k_x^{'}$ also tends to zero. Let us start from simplification summation function (\ref{eq:hyperbolic sum}):
\begin{equation}
\label{eq:app sum hyperbolical zero limit}
\begin{gathered}
    S_{\text{hyp}}(k_x^{'}, N_x) = \lim_{k_x^{'} \rightarrow 0} \frac{1}{4} \left(\frac{\sinh{k_x^{'} (2 N_x + 1)}}{\sinh{k_x^{'}}} - (2 N_x + 1) \right) \\
    = \frac{1}{6} N_x (N_x + 1) (2 N_x + 1) (k_x^{'})^2.    
\end{gathered}
\end{equation}
The wave functions (first of Eq. (\ref{eq:zGNR localized wave function final})) can be written as
\begin{equation}
\label{eq:app zGNR k_y^c wave function}
\begin{gathered}
    \mqty(\psi_{A,\textbf{i}} \\ \psi_{B,\textbf{i}} ) = \lim_{\eta, k_x^{'} \rightarrow 0} \frac{(-1)^n e^{i (k_y^c + \eta) m}}{\sqrt{N_y} k_x^{'} \sqrt{N_x (N_x + 1) (2 N_x + 1)/6}} \\
    \cdot \mqty(s k_x^ {'}  n     \\      k_x^ {'}  (N_x-n)) \\
    = \frac{(-1)^n \sqrt{6} e^{i k_y^c m}}{\sqrt{N_y N_x (N_x + 1) (2 N_x + 1)}} \mqty(s \cdot n  \\  N_x-n).  
\end{gathered}
\end{equation}
This wave functions differ from (\ref{eq:zGNR k_y^c case}) by the factor $(-1)^{N_x + 1}$, but it can be easily absorbed by renormalization. The energy related function $\epsilon (k_x, k_y)$ for this case has the same values as in (\ref{eq:k_y^c energy}).

\subsection{Left limit of wave functions at the point $k_y = \pi - k_y^c$} \label{app:Left limit of wave functions at pi - k_y^c}
We need to use complex plane continuation of the form $k_x = 0 \pm i k_x^{'}$ for the left limit $k_y = \pi - k_y^c - \eta$ ($\eta \rightarrow 0^+$) leads to limiting to zero of $k_x^{'}$. Summation function (\ref{eq:hyperbolic sum}) will have form (\ref{eq:app sum hyperbolical zero limit}), so the wave functions (second of Eq. (\ref{eq:zGNR localized wave function final})) can be simplified as follows
\begin{equation}
\label{eq:app zGNR pi-k_y^c wave function}
\begin{gathered}
    \mqty(\psi_{A,\textbf{i}} \\ \psi_{B,\textbf{i}} ) = \lim_{\eta, k_x^{'} \rightarrow 0} \frac{e^{i (\pi - k_y^c - \eta) m}}{\sqrt{N_y} k_x^{'} \sqrt{N_x (N_x + 1) (2 N_x + 1)/6}} \\
    \cdot \mqty(s k_x^ {'}  n     \\      k_x^ {'}  (N_x-n)) \\
    = \frac{(-1)^m \sqrt{6} e^{- i k_y^c m}}{\sqrt{N_y N_x (N_x + 1) (2 N_x + 1)}} \mqty(s \cdot n  \\  N_x-n). 
\end{gathered}
\end{equation}

Let us calculate the function $\epsilon(k_x,k_y)$ (second of Eq. (\ref{eq:special case energies})):
\begin{equation}
\label{eq:app energy pi - k_y^c}
\begin{gathered}
    \epsilon (0, \pi - k_y^c) = \lim_{\eta, k_x^{'} \rightarrow 0} \left( 3 + 4 \cosh{k_x^ {'}} \cos{(\pi - k_y^c - \eta)} \right. \\
    \left. + 2 \cos{2 (\pi - k_y^c - \eta)} \right)^{1/2} = \sqrt{3 - 4 \cos{k_y^c} + 2 \cos{2 k_y^c}} \\
    = \frac{1}{N_x + 1},
\end{gathered}
\end{equation}
which is identical to the value of this function at another transition point $k_y = k_y^c$ (\ref{eq:k_y^c energy}).

\subsection{Right limit of wave functions at the point $k_y = \pi - k_y^c$} \label{app:Right limit of wave functions at pi - k_y^c}
We have extended states (\ref{eq:zGNR extended wave function final}) in the region $k_y = \pi - k_y^c + \eta$ ($\eta > 0$). The wavenumber $k_x$ tends to zero when $\eta \rightarrow 0$. The summation function (\ref{eq:sum sin^2}) in this case can be calculated similarly to (\ref{eq: sum sin^2 transition point}):
\begin{equation}
\label{eq: sum sin^2 transition point 2}
\begin{gathered}
    S(k_x, N_x) = \lim_{k_x \rightarrow 0} \sum_{n=1}^{N_x} \sin^2 k_x n = k_x^2 \sum_{n=1}^{N_x} n^2 \\
    = k_x^2 \frac{N_x(N_x+1)(2N_x + 1)}{6}.
\end{gathered}
\end{equation}
The sign function $s_1$ (\ref{eq:s1}) also transforms:
\begin{equation}
    s_1 = \lim_{k_x \rightarrow 0} \text{sign} \left(\sin k_x (N_x+1) \right) = 1.
\end{equation}
Finally, we can write simplified wave functions
\begin{equation}
\label{eq:app zGNR pi-k_y^c wave function 2}
\begin{gathered}
    \mqty(\psi_{A,\textbf{i}} \\ \psi_{B,\textbf{i}} ) = \lim_{\eta, k_x \rightarrow 0} \frac{e^{i (\pi - k_y^c + \eta) m}}{\sqrt{N_y} k_x \sqrt{N_x(N_x+1)(2N_x + 1)/6}} \\
    \cdot \mqty(s k_x n  \\  k_x (N_x-n)) \\
    = \frac{(-1)^m \sqrt{6} e^{-i k_y^c m}}{\sqrt{N_y N_x(N_x+1)(2N_x + 1)}} \mqty(s \cdot n  \\ N_x-n).    
\end{gathered}
\end{equation}
This result coincides with the left limit at the point $k_y = \pi - k_y^c$ (\ref{eq:app zGNR pi-k_y^c wave function}). The function $\epsilon(k_x, k_y)$ at this point has value (\ref{eq:app energy pi - k_y^c}).

In all considered cases left and right limits give identical (up to renormalization) wave functions and coinciding energies. This result is in accordance with the principle of continuity.

\section{Proof that pairs $(k_x,k_y)$ and $(\pi - k_x, \pi - k_y)$ describe identical states} \label{app:finite sample proof of change}

\subsection{Extended states}
We start from proving that the function $\epsilon(k_x, k_y)$ has the same values for these pairs of $(k_x,k_y)$:
\begin{equation}
\begin{gathered}
    \epsilon (\pi - k_x, \pi - k_y) = \\
    = \sqrt{3 + 4 \cos{(\pi - k_x)} \cos{(\pi - k_y)} + 2 \cos{2 (\pi - k_y)}} \\
    = \sqrt{3 + 4 \cos{k_x} \cos{k_y} + 2 \cos{2 k_y}} = \epsilon (k_x, k_y).
\end{gathered}
\end{equation}
This means that these states also have identical energies.

Now let us do the same thing for wave functions (\ref{eq:extended states finite system}). First, we study what happens to the sign function $s_1$ (\ref{eq:s1}) which depends only on $k_x$:
\begin{equation}
\begin{gathered}
    s_1 (\pi - k_x) = \text{sign} \left(\sin ((\pi - k_x) (N_x+1)) \right) \\
    = \text{sign} \left(- (-1)^{N_x + 1} \sin (k_x (N_x+1)) \right) \\
    = (-1)^{N_x} \text{sign} \left(\sin (k_x (N_x+1)) \right) = (-1)^{N_x} s_1(k_x).
\end{gathered}
\end{equation}
The summation function $S(k,N)$ also does not change for $k \rightarrow \pi - k$. The second step is to rewrite the wave functions (\ref{eq:extended states finite system}):
\begin{equation}
\begin{gathered}
    \mqty(\psi_{A,\textbf{i}} \\ \psi_{B,\textbf{i}} ) = \frac{\sin{(\pi - k_y) m}}{\sqrt{S(k_x, N_x) S(k_y, N_y)}} \\
    \cdot \mqty((-1)^{N_x} s_1(k_x) s \sin ((\pi - k_x) n)  \\    \sin ((\pi - k_x) (N_x-n))) \\
    = \frac{(-1)^{m + 1} \sin{k_y m}}{\sqrt{S(k_x, N_x) S(k_y, N_y)}} \mqty((-1)^{N_x + n + 1} s_1(k_x) s \sin (k_x n)  \\  (-1)^{N_x - n + 1}  \sin (k_x (N_x-n))) \\
    = \frac{(-1)^{N_x + m + n} \sin{k_y m}}{\sqrt{S(k_x, N_x) S(k_y, N_y)}} \mqty(s_1(k_x) s \sin (k_x n)  \\  (-1)^{- 2n}  \sin (k_x (N_x-n))).
\end{gathered}
\end{equation}
These wave functions coincide with wave functions for $(k_x,k_y)$ (\ref{eq:extended states finite system}), because $(-1)^{2n} = 1$ and $(-1)^{m + n}$ are constant for all sites (it depends on cells numeration choice: in our case of numeration like in Fig. \ref{fig: armchair nanoribbon} it is always $+1$) and finally $(-1)^{N_x}$ can be absorbed by renormalization.

\subsection{Localized states}
For the region $k_y \in \left( k_y^c ; \pi/2 \right)$ wave functions and eigenenergies have the form (\ref{eq:localized wave function finite system}) and (\ref{eq:energies for finite system in localized state}) respectively. If one wants to make the change $(k_x,k_y) \rightarrow (\pi - k_x, \pi - k_y)$ they need to know the wave functions and energies for the region $k_y \in \left(\pi/2; \pi - k_y^c \right)$. We derive wave functions and eigenenegies for $k_y \in \left(\pi/2; \pi - k_y^c \right)$ from wave functions (second of (\ref{eq:zGNR localized wave function final})) and eigenenergies (second of (\ref{eq:special case energies})) for zigzag nanoribbons:
\begin{equation}
\label{app:localized wave function finite system ky > pi/2}
    \mqty(\psi_{A,\textbf{i}} \\ \psi_{B,\textbf{i}} ) = \frac{\sin{k_y m}}{\sqrt{S_{\text{hyp}}(k_x^{'}, N_x) S(k_y, N_y)}} \mqty(s \sinh (k_x^ {'}  n)     \\      \sinh (k_x^ {'}  (N_x-n)) ),
\end{equation}
\begin{equation}
\label{app:energies for finite system in localized state ky > pi/2}
    E_s = s \cdot t \sqrt{3 + 4 \cosh{k_x^ {'}} \cos{k_y} + 2 \cos{2 k_y}}.
\end{equation}
Now, we make the replacement $(k_x,k_y) \rightarrow (\pi - k_x, \pi - k_y)$ in (\ref{app:localized wave function finite system ky > pi/2}) and (\ref{app:energies for finite system in localized state ky > pi/2}) to compare these results with (\ref{eq:localized wave function finite system}) and (\ref{eq:energies for finite system in localized state}) respectively.

Let us start by comparing energies. The parameter $k_x^ {'}$ does not change because the change $k_x \rightarrow \pi - k_x$ is already included in different types of complex plane continuations (\ref{eq:complex plane continuation}). So we only need to change $k_y \rightarrow \pi - k_y$ in (\ref{app:energies for finite system in localized state ky > pi/2}):
\begin{equation}
\begin{gathered}
    E_s = s \cdot t \sqrt{3 + 4 \cosh{k_x^ {'}} \cos{(\pi - k_y)} + 2 \cos{2 (\pi - k_y)}} \\
    = s \cdot t \sqrt{3 - 4 \cosh{k_x^ {'}} \cos{k_y} + 2 \cos{2 k_y}},
\end{gathered}
\end{equation}
which coincides with energy (\ref{eq:energies for finite system in localized state}).

Now we compare wave functions in the same way, we exchange only $k_y \rightarrow \pi - k_y$ (summation functions remain the same):
\begin{equation} \label{app:localized wave function finite system ky > pi/2 exchange}
\begin{gathered}
    \mqty(\psi_{A,\textbf{i}} \\ \psi_{B,\textbf{i}} ) = \frac{\sin{(\pi - k_y) m}}{\sqrt{S_{\text{hyp}}(k_x^{'}, N_x) S(k_y, N_y)}} \mqty(s \sinh (k_x^ {'}  n)     \\      \sinh (k_x^ {'}  (N_x-n)) ) \\
    = \frac{(-1)^{m + 1} \sin{k_y m}}{\sqrt{S_{\text{hyp}}(k_x^{'}, N_x) S(k_y, N_y)}} \mqty(s \sinh (k_x^ {'}  n)     \\      \sinh (k_x^ {'}  (N_x-n)) ).
\end{gathered}
\end{equation}
This result is identical to wave functions (\ref{eq:localized wave function finite system}), because factors $(-1)^m$ in (\ref{app:localized wave function finite system ky > pi/2 exchange}) and $(-1)^n$ in (\ref{eq:localized wave function finite system}) work the same way (explained at the end of Sec. \ref{sec:wave functions for zigzag nanoribbons}) and $(-1)$ can be absorbed by renormalization.
\end{document}